\documentclass{icrc2009}

\usepackage{graphicx}   
\usepackage{caption}    
\usepackage[font=footnotesize]{subfig} 
\usepackage{fixltx2e}
\usepackage{url}

\newcommand{\shorttitle}[1]%
{\markboth{Proceedings of the 31\MakeLowercase{$^{st}$} ICRC, {\L}\'{o}d\'{z} 2009}{#1} }
\newcommand{\etal}{\MakeLowercase{\textit{et al. }}} 

\begin{document}
\title{Studies of the Influence of Moonlight on Observations with the MAGIC Telescope}

\author{\IEEEauthorblockN{Daniel Britzger\IEEEauthorrefmark{1}\IEEEauthorrefmark{6},
		             	  Emiliano Carmona\IEEEauthorrefmark{1},
                         Pratik Majumdar\IEEEauthorrefmark{2},
                          Oscar Blanch\IEEEauthorrefmark{3},\\
                          Javier Rico\IEEEauthorrefmark{4}\IEEEauthorrefmark{3},
                          Julian Sitarek\IEEEauthorrefmark{1}\IEEEauthorrefmark{5} and
                          Robert Wagner\IEEEauthorrefmark{1} for the MAGIC Collaboration}\\
\IEEEauthorblockA{\IEEEauthorrefmark{1}Max-Planck-Institut f\"{u}r Physik, D-80805 M\"{u}nchen, Germany}
\IEEEauthorblockA{\IEEEauthorrefmark{2}Deutsches Elektronen-Synchrotron (DESY), D-15738 Zeuthen, Germany}
\IEEEauthorblockA{\IEEEauthorrefmark{3}IFAE, Edifici Cn., Campus UAB, E-08193 Bellaterra, Spain}
\IEEEauthorblockA{\IEEEauthorrefmark{4}ICREA, E-08010 Barcelona, Spain }
\IEEEauthorblockA{\IEEEauthorrefmark{5}University of {\L}\'{o}d\'{z}, PL-90236 {\L}\'{o}d\'{z}, Poland}
\IEEEauthorblockA{\IEEEauthorrefmark{6}britzger@mpp.mpg.de}
}
\shorttitle{Britzger~\etal, Observations~under~moonlight with~MAGIC}
\maketitle
\begin{abstract}
The ground-based imaging atmospheric Cherenkov technique is currently the most
powerful observation method for very high energy gamma rays. With its specially
designed camera and readout system, the MAGIC Telescope is capable of observing
also during nights with a comparatively high level of night-sky background
light.
This allows to extend the MAGIC duty cycle by 30$\%$ compared to
dark-night observations without moon.
Here we investigate the impact of increased background light on single-pixel
level and show the performance of observations in the presence of moonlight
conditions to be consistent with dark night observations.
 \end{abstract}
\begin{IEEEkeywords}
Imaging Atmospheric Cherenkov Technique, MAGIC telescope, moonlight
\end{IEEEkeywords}
\section{Introduction}
 MAGIC-I is an Imaging Atmospheric Cherenkov Telescope (IACT) located on the
Canary Island of La Palma at 2200m a.s.l. \cite{MAGIC}. The IACT technique is a very
successful observation method for very high energy cosmic-ray particles, particularly
gamma rays with energies ranging from $\sim$~100~GeV up to some 10~TeV.
 The underlying background from single photons from the night sky (night-sky background; NSB) as measured with MAGIC-I is at 0.12~phe/ns/pixel (photoelectrons per nanosecond per 0.1$^\circ$ diameter photomultiplier [PMT] pixel) for an extragalactic field of view and around 0.18~phe/ns/pixel for a galactic area.
 Here we define an \emph{NSB} unit as the background level which MAGIC-I records while pointing to a galactic celestial field comparable to the Crab Nebula region. It is directly proportional to 0.18~phe/ns/pixel and to the anode current at $\sim$1.0$\mu$A, or with a square-root proportion to the pedestal RMS at 0.8~phe/pixel/event.\\
Under moonlight (or twilight) conditions, which seem still feasible for observations with MAGIC, the background increases by up to a factor of 6, i.e., to above 1.0~phe/ns/pixel.\\
A first study had been performed for the initial MAGIC-I telescope
setup with 300 MHz readout \cite{moonpaperAstroPH}.  It could be shown
that the noise induced by the moonlight itself contributes only in a
negligible way. The distributions of the Hillas parameters (parameters
that describe the air shower images \cite{Hillas}) are in good
agreement with that of dark night observations. This makes the
analysis of data taken under moonlight conditions straightforward, as
no special treatment is required for calibration. The energy threshold
of the telescope was shown to be only marginally affected by increased
light levels due to moonlight. Since the initial study
\cite{moonpaperAstroPH}, several substantial improvements in the hardware and
analysis were implemented. The readout system was upgraded to a 2-GSamples/s
Flash-ADC system \cite{Goebel} along with new signal receivers,
allowing for a substantially improved air shower image cleaning that
now also takes the air shower timing into account
\cite{TimingIC}, reducing the NSB included in the pixel
signal. The study presented in this paper is based on data taken
with the new readout and subsequent analysis refinements, but also
accounts for the now-standard observation mode in MAGIC, the
\emph{wobble} mode \cite{Daum}, which allows to record
on-source and off-source (background control) data simultaneously. In this paper we
analyze a large sample of Crab Nebula data taken under vastly
different moonlight conditions with the above mentioned system. As
done in \cite{moonpaperAstroPH}, we focus on the relevant physics
analysis quantities -- the resulting differential energy spectrum and
flux of the Crab Nebula as well as the achieved sensitivity for a
Crab-like source. We investigate the telescope performance with the
MAGIC standard analysis \cite{Moralejo}, employing two
exemplarily chosen image cleanings (see below). The results are
compared with expectations from Monte Carlo (MC) simulations.
%
%
\section{Considerations for observations under moonlight}
The MAGIC-I camera was designed to allow for observations under moonlight conditions. It consists of an inner hexagon with 396 PMTs with a field of view (FOV) of $0.1^\circ$ each, and an outer region with 180 PMTs of $0.2^\circ$ FOV. The PMTs have a low gain amplification of $3\times10^4$.
This prevents the last PMT dynode from too high damage by frequent pulses from diffuse background light, like moonlight.\\
The illumination from moonlight depends on the lunar phase, the distance of the moon from earth, the altitude of the moon, and the separation angle between the observed source and the moon. For IACT, particularly small lunar phases, low lunar altitudes and separation angles between 25$^\circ$-110$^\circ$ are relevant: within 80\% of moontime observations the altitude of the moon does not exceed 30$^\circ$. The first/last-quarter moon has only $\sim$10\% of the full-moon brightness~\cite{Allen}.
For estimating the brightness of moonlight in the MAGIC-I telescope camera, a model has been developed \cite{DipDani} similar to~\cite{KSmodel,Garstang}, which holds for those special conditions. It incorporates differences to previous predictions (e.g.~\cite{Allen,KSmodel,AustinMethod}), particularly for early lunar phases and small moon altitudes. It is presented in \cite{DipDani} in detail and can be used for automated MC simulations as well as for observation scheduling.\\
The main motivation for observing under moonlight (or twilight) is the gain in observation time. This is particularly relevant for long-term source monitoring, for the coverage of multiwavelength observations, or to increase the probability to detect gamma-ray bursts and to follow-up flares, e.g., in active galactic nuclei.
MAGIC has a dark night observation time of 1600~h per year, dark time being defined as astronomical night (sun below -18$^\circ$ altitude) and the moon being below horizon.
Conservatively assuming a twice higher NSB, additional 300~h per year are accessible. This assumption corresponds to a 30\% illuminated moon and a separation angle above 50$^\circ$.
More ambitiously allowing for an increased background of 6$\times$NSB extends the observation time by 550~h per year. This is equivalent to a 70\% illuminated moon at a separation angle more than 50$^\circ$.
\section{Observations performed under moonlight}
\begin{figure}[!t]
 \centering
 \includegraphics[width=\linewidth]{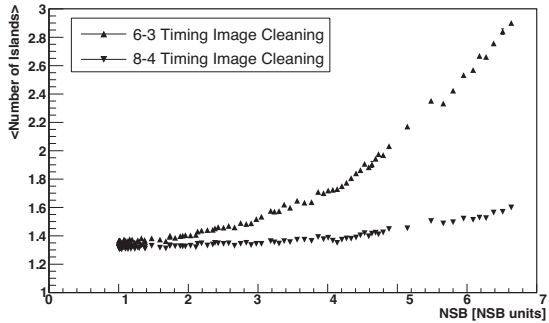}
 \caption{Mean number of islands depending on NSB (and \texttt{Size}
   $>$ 100~phe) for two different time image cleanings applied to the
   same data.}
 \label{NoIsl}
\end{figure}

 \begin{figure}[!t]
 \centering
 \includegraphics[width=\linewidth]{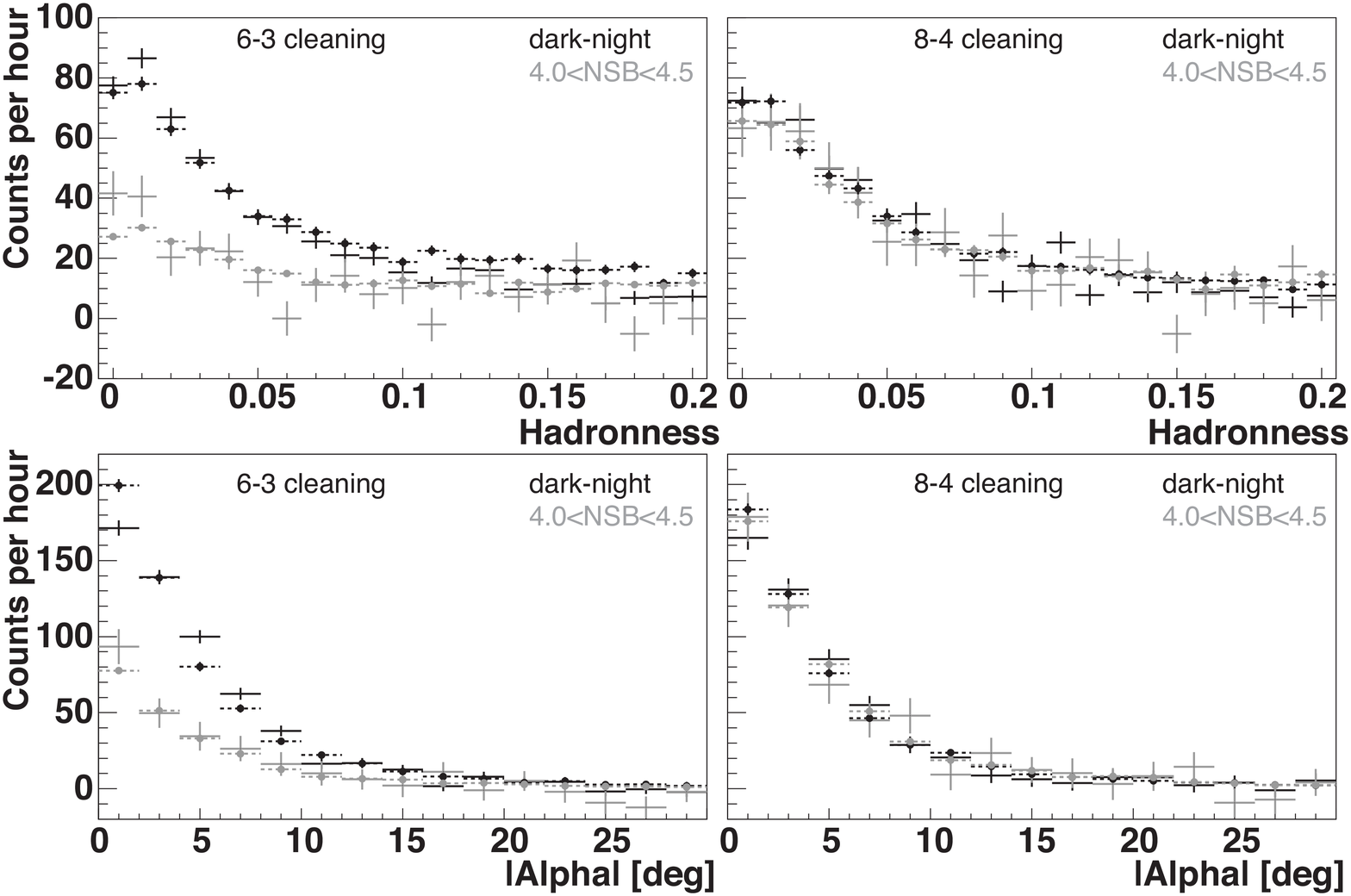}
 \caption{Hadronness-excess (upper plots) and \texttt{|Alpha|}-excess (lower plots) for two different cleaning parameters (left: \emph{6-3}; right: \emph{8-4}). Black markers indicate dark night data, while the grey markers are for 4.0 $<$ NSB $<$ 4.5. Dotted markers are corresponding MC simulations. A size cut of 220~phe was applied.}
 \label{Hadronness}
\end{figure}

More than 42 hours of Crab Nebula data were recorded under moonlight
conditions. After the hardware upgrade of the DAQ system, dedicated
data was taken within 27 days from February 2007 to February 2008
under very different moon phases.  The moon phase ranges up to 53\% at
55$^\circ$ altitude and to 80\% at low altitudes, respectively. The separation
angle ranges from 25$^\circ$ to 130$^\circ$. The Crab Nebula zenith
angle was between 5$^\circ$ and 30$^\circ$.
The moonlight increases the background from 1 up to 6.5 NSB in the
data. The level-0 trigger thresholds for each pixel were adjusted with
an automatic individual pixel control system as to achieve an almost
constant trigger rate. (Splitters divide the signal before the trigger
level, so the recorded images are not affected by these changing trigger
thresholds.) We record the anode direct currents (DC) for each inner PMT
using an ADC with an integration window of a few $\mu$s sampled with 3
Hz frequency. The median and mean values of the DCs are
calculated in bins of one minute size. The DCs are therefore
 proportional to the NSB.

\section{The impact of moonlight on MAGIC observations}
The basic problem of observations under moonlight conditions is given
by the increased background fluctuations in each individual pixel.
During moonlight observations, the recorded phe rate rises to
$<geq$~1.0~phe/ns/pixel. A 300~GeV gamma-ray induced shower produces
a signal of 2--20~phe/pixel in $\sim$30 pixels, typically arriving
within 3~ns; a resulting pedestal RMS of 3.2~phe/pixel/event is
therefore comparatively high.\\ Accidental triggers from those
fluctuations are suppressed by the automatically increased
discriminator thresholds and are removed due to their small
sizes. However, when ``cleaning'' the images, so-called islands remain
as artifacts in the air shower images. An island is defined as a group
of pixels detached from another group after image cleaning. This
feature affects the analysis of the shower images.\\ Here we use the
MAGIC standard analysis \cite{Moralejo,CrabPaper}, which additionally
applies an image cleaning with timing constraints as presented
in~\cite{TimingIC}. The standard image cleaning for non-moon
observations (referred to as \emph{6-3}) keeps core pixels with
charge $>$6 phe and a maximum spread of 4.5~ns plus boundary pixels
with charge $>$3 phe and a maximum of 1.5~ns delay from the
neighboring core pixel. Pixels not fulfilling these constraints are ignored in
the image parametrization.  With increasing NSB, the mean number of
islands increases dramatically~(Fig.~\ref{NoIsl}).
Besides the standard cleaning levels 6-3, we
investigated exemplarily a higher cleaning level, applying
the same timing constraints, but requires 8~phe for core
pixels and 4~phe for boundary pixels (called~\emph{8-4}
cleaning).\\ The choice of image cleaning parameters
leaves its traces throughout the analysis chain.
The mean number of islands
can be used as a quality parameter, in the sense that with increasing NSB the
mean number of islands stays constant until a certain NSB level, while it increases
for higher background (cf. Fig.~\ref{NoIsl}).  We recognize a significant
deviation from the dark night level at $2.0\times$NSB for the
\emph{6-3} cleaning, and around $4.5\times$NSB for the \emph{8-4}
cleaning. In the further analysis, we apply a cut on the number of islands $\leq$ 2.
%
%
%

 \begin{figure}[!t]
 \centering
  \includegraphics[width=3.0in]{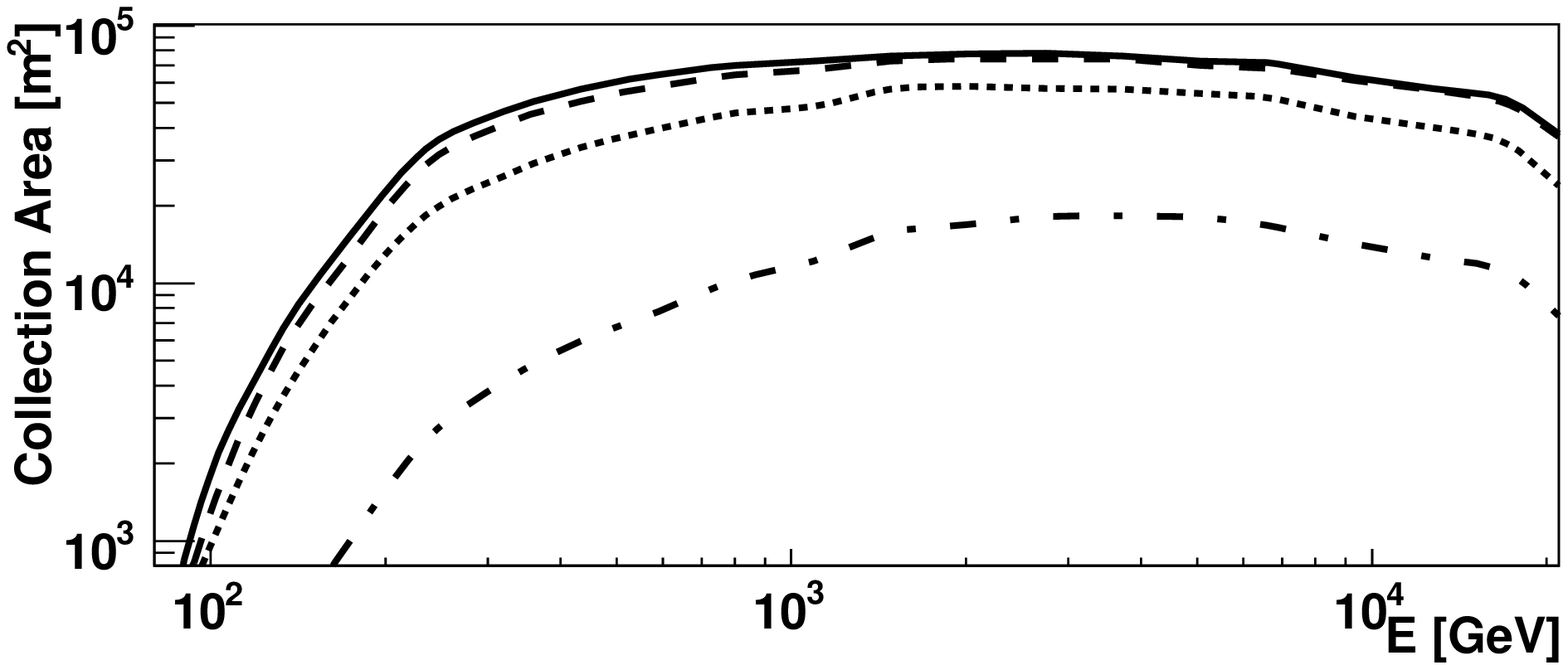}
  \includegraphics[width=3.0in]{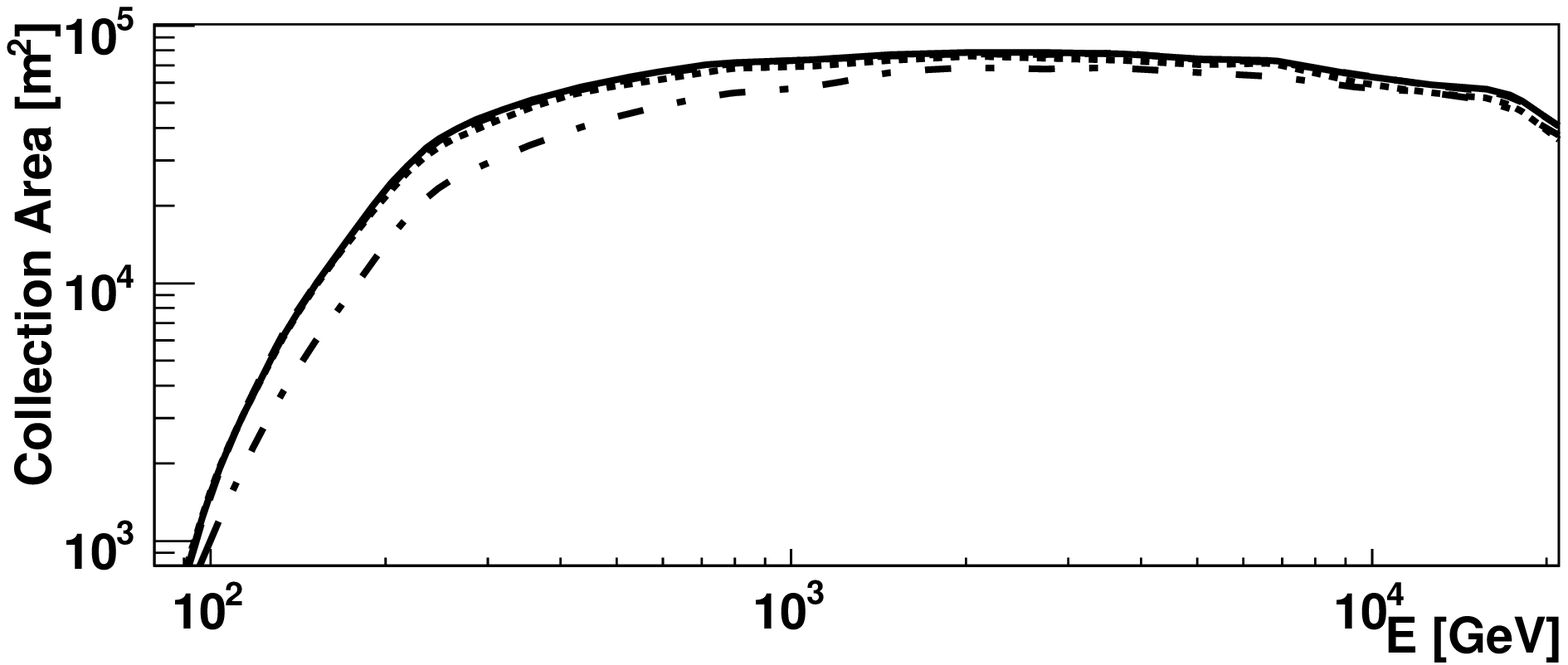}
  \caption{Effective collection area after cuts for (a) \emph{6-3} and (b)
\emph{8-4} timing image cleaning for four simulated background levels.
Solid lines represent for dark night MCs (1.0 NSB), dashed lines correspond to 2.5
NSB, dotted lines to 3.75 NSB, and dash-dotted lines to 5.5 NSB.}
 \label{AEff}
\end{figure}
\begin{figure*}[!t]
  \centerline{
   {\includegraphics[width=3.1in]{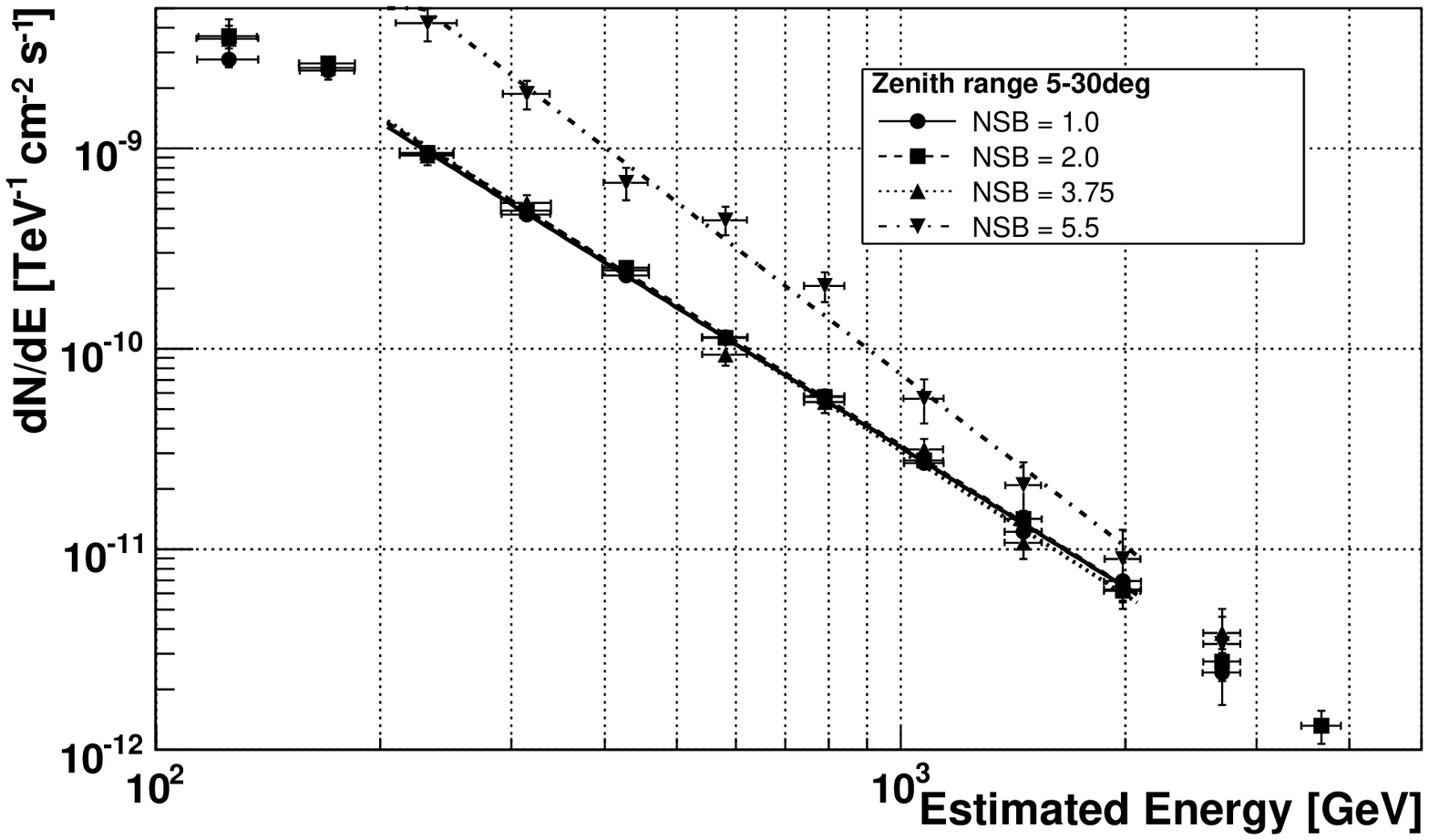} \label{DiffSp6-3}}
   \hfill
   {\includegraphics[width=3.1in]{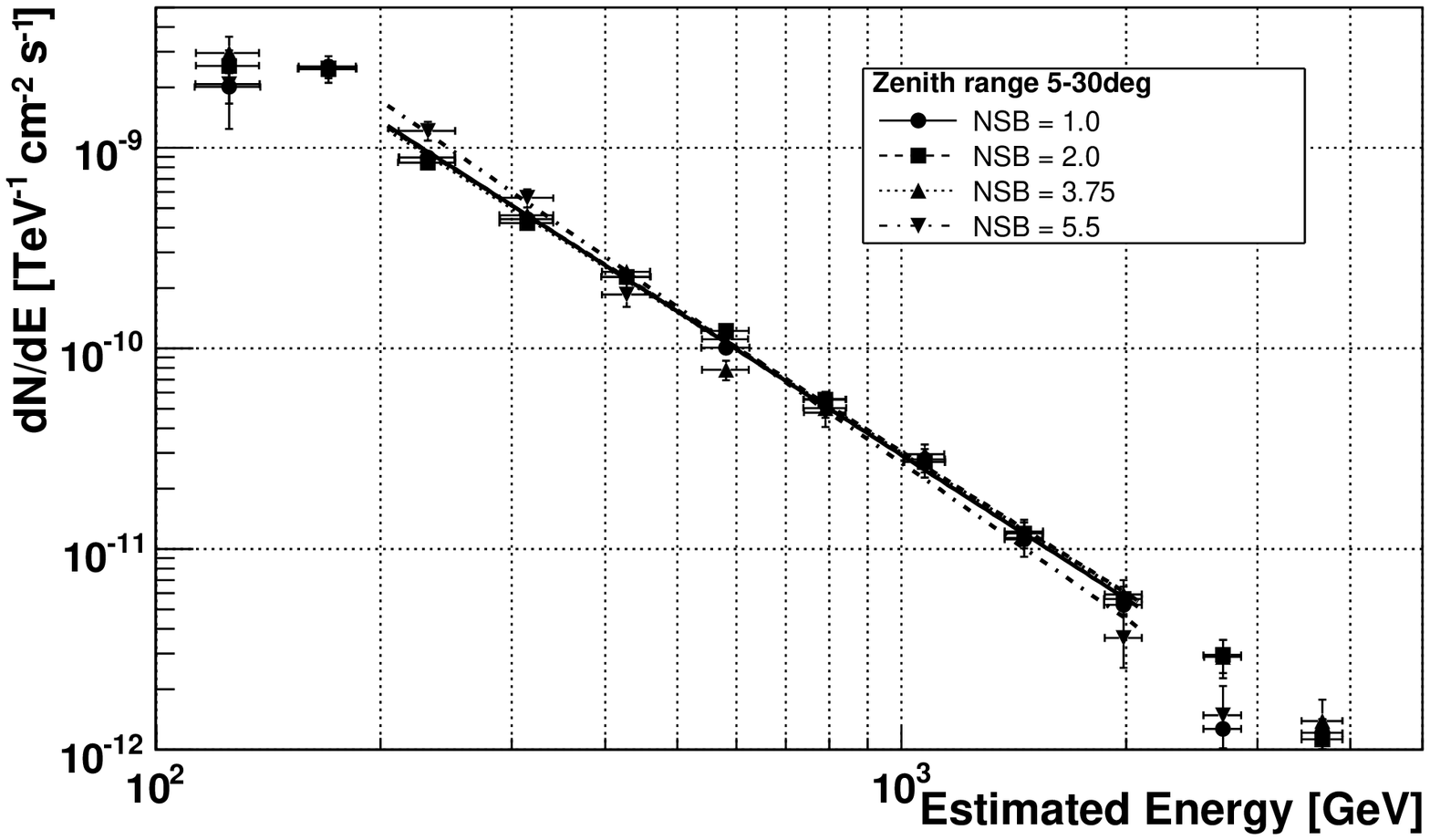} \label{DiffSp8-4}}
            }
  \caption{Differential energy spectrum of Crab Nebula (left: 6-3, right: 8-4
image cleaning) for four different moonlight illuminations. The continuous line
represents to the dark night spectrum derived from $t_{\rm eff}$=26~h effective
observation time. The dashed curve is derived for 2.0$<$NSB$<$3.0
$t_{\rm eff}=$7.5~h), the dotted curve for 3.0$<$NSB$<$4.5
($t_{\rm eff}=$3.8~h), and the dash-dotted line for 4.5$<$NSB$<$6.0
($t_{\rm eff}=$2.3~h).
}
  \label{DiffSpectrum}
\end{figure*}

  Within the MAGIC standard analysis the discrimination of hadronic background
events (``$\gamma$/hadron separation'') is performed using a decision tree
algorithm \cite{RF}. For training of the \emph{random forest}, we used
dark-night MC simulations for $\gamma$-like events and real data for
hadron-like
events. Two individual \emph{random forests} were
generated for the different cleanings. The same decision trees are used for the complete data set, i.e.,
also for bright moonlight data, since the Hillas parameters~\cite{Hillas} were
found to be constant on average during moonlight
observations~(see~e.g~\cite{moonpaperAstroPH}).\\
  Fig.~\ref{Hadronness} illustrates the effect of moonlight on the $\gamma$/hadron separation for the investigated data set. During dark night observations (black counts) the hadronness-excess (classification parameter of the decision tree) indicates high excess for gamma-like particles.
  While for NSB between 4.0 and 4.5 (grey counts) the separation capability for
\emph{8-4} cleaned data is almost dark-like, there is a decreased hadronness
excess for \emph{6-3} cleaned data (upper-left plot).
  The filled circles represent the hadronness of MC simulations
  (black: dark like, grey: 3.75$\times$ NSB). Those are in good
  agreement with the data.
  \\
The \texttt{|ALPHA|}-excess plots are mainly affected by the decreased $\gamma$/hadron-separation power under strong moon conditions as also Fig.~\ref{Hadronness} indicates, since only events with hadronness $<$ 0.1 are taken into account. Therefore, the \texttt{|ALPHA|}-excess plot for \emph{6-3} cleaning shows a less significant peak under high NSB levels. Again, the MC simulations are in good agreement with the data.\\
 For        energy reconstruction, also performed using a
 \emph{random forest} method \cite{RF}, we employed again only the ``dark
 night'' decision tree for all NSB levels. We checked that this results in
 correct energy estimates, showing again that the quality of
 the images is not affected by the increased NSB. Also the analysis energy threshold was checked to be constant at $\sim$190~GeV for the applied size cut of 220~phe for all NSB levels.
\subsection{Sensitivity}

 \begin{figure}[!t]
 \centering
 \includegraphics[width=3.2in]{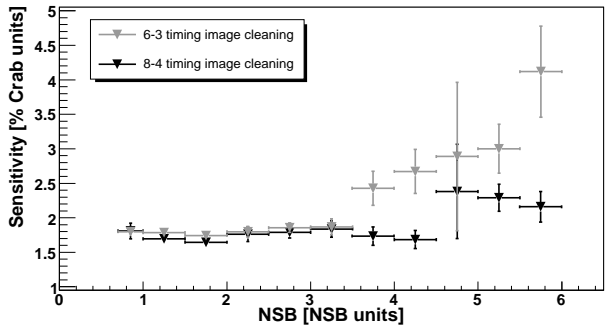}
 \caption{Sensitivity in \emph{Crab units} of MAGIC-I for the standard analysis (grey) and for an exemplarily applied \emph{8-4} timing image cleaning depending on the NSB. }
 \label{Sensitivity}
\end{figure}

 We apply cuts in Hadronness ($<$ 0.05), size ($>$400 phe), and
 \texttt{|ALPHA|} ($<$ 6$^\circ$) and calculate the \texttt{ON} and
 \texttt{OFF}-events. The resulting sensitivity of MAGIC-I above an
 energy of 250~GeV for
 different NSB levels is presented in Fig.~\ref{Sensitivity} in Crab
 units. The sensitivity is computed applying the standard MAGIC
 analysis and using the standard dark-night MC.  We conclude that
 the sensitivity remains constant up to 3.5 NSB for the MAGIC standard
 6-3 cleaning (grey counts). If higher cleaning parameters are
 applied, the sensitivity is constant for even higher NSB.

\subsection{Spectrum under moonlight conditions}
We performed MC simulations for four different
NSB levels: 1.0, 2.5, 3.75, and 5.5 NSB.
We crosschecked that the resulting pedestal RMS of the pixel
calibration is the same as in real data. The simulated trigger
threshold (DT) was also adjusted in MC with values from
data.\\ We apply cuts on Hadronness $<$ 0.1, size $<$ 220 phe and
\texttt{|ALPHA|} $<$ 8$^\circ$. Fig.~\ref{AEff} shows the resulting
effective collection area ($A_{\rm eff}$). While for \emph{6-3} cleaning
parameters, $A_{\rm eff}$ already decreases around NSB $>$ 2.0, it is
still compatible with dark night until 3.75 NSB for \emph{8-4}. This
shows that the dark-night MC can be used with the MAGIC standard
analysis (\emph{6-3} cleaning) is stable until $\sim$2.5$\times$NSB and
$\sim$3.75$\times$NSB with the \emph{8-4} cleaning.

We used the estimated $A_{\rm eff}$ from Fig.~\ref{AEff} to obtain
NSB-corrected differential energy spectra for Crab under different
moonlight conditions.  The resulting spectra for four different NSB
levels are shown in Fig.~\ref{DiffSpectrum}. An exponential spectrum
is assumed and fitted from 200~GeV to 2~TeV.
%
For the 6-3
cleaning, the Crab spectrum remains stable up to 2.5 NSB and it can be
recovered by using $A_{\rm eff}$ computed from MC with
the corresponding NSB up to about 4.0 NSB.
For the 8-4 cleaning, no correction is needed up to even 6.0
NSB; With an appropriate correction one could control even higher NSB levels.

\section{Conclusions}
We performed a study of the effect of moonlight on MAGIC observations
after the changes introduced in the DAQ, software and observation
mode during the past years. As shown in previous studies \cite{moonpaperAstroPH}, data
quality is not affected by increased NSB levels. For observations
under moonlight the MAGIC standard analysis ({\emph 6-3} cleaning) is
robust up to 2.5 NSB.
Using dedicated
moonlight simulations, the dark-like spectrum and flux can be
recovered for even higher NSB levels.\\
\newpage
 If a higher image
cleaning is applied, observations under more intense moonlight are
possible with MAGIC. The investigated 8-4 timing image cleaning is
robust until 4.5 times the galactic NSB with dark night Monte Carlo
simulations. This is comparable to a first/third-quarter moon (50\%
illuminated; altitude on La Palma $<$ 60$^\circ$) at a small separation angle
of 30$^\circ$.
Dedicated moonlight simulations give the
possibility to perform a correct analysis even for higher NSB
levels.\\ The sensitivity of MAGIC-I is darktime-like up to 3.5 times
increased NSB.\\
For a moon illumination of 70\%, and assuming a source can be found at a separation angle $>$50 degrees, the NSB is below 6~NSB. This allows us to add about 550 hours (additional 30\%) of observation time per year, without any significant change in the analysis.

\end{document}